\def\ref{\par\noindent\hang}

\def\spose#1{\hbox to 0pt{#1\hss}}
\def\approxlt{\mathrel{\spose{\lower 3pt\hbox{$\sim$}}
	\raise 2.0pt\hbox{$<$}}}
\def\approxgt{\mathrel{\spose{\lower 3pt\hbox{$\sim$}}
	\raise 2.0pt\hbox{$>$}}}

\def\multleft#1{\hbox to size{\vbox {\halign {\lft{##}\cr #1}}\hfill}\par}
\def\multright#1{\hbox to size{\vbox {\halign {\rt{##}\cr #1}}\hfill}\par}

\def\degmark{^\circ}
\def\today{\ifcase\month\or January\or February\or March\or April\or May\or
      June\or July\or August\or September\or October\or November\or December\fi
      \space\number\day, \number\year}
\def\$<${\thinspace}
\def\s{\hbox{\phantom{5}}}	%one space
\def\boxit#1{\vbox{\hrule\hbox{\vrule\kern3pt\vbox{\kern3pt
          #1 \kern3pt}\kern3pt\vrule}\hrule}}

\def\cm{{\rm\thinspace cm}}
\def\dyn{{\rm\thinspace dyn}}
\def\erg{{\rm\thinspace erg}}

\def\MeV{{\rm\thinspace MeV}}

\def\keV{{\rm\thinspace keV}}
\def\km{{\rm\thinspace km}}
\def\kpc{{\rm\thinspace kpc}}

\def\Mpc{{\rm\thinspace Mpc}}
\def\Msun{\hbox{$\rm\thinspace M_{\odot}$}}
\def\pc{{\rm\thinspace pc}}

\def\s{{\rm\thinspace s}}
\def\yr{{\rm\thinspace yr}}

\def\Hz{{\rm\thinspace Hz}}
\def\MHz{{\rm\thinspace MHz}}
\def\GHz{{\rm\thinspace GHz}}

\def\pcmcu{\hbox{$\cm^{-3}\,$}}

\def\dynpcmsq{\hbox{$\dyn\cm^{-2}\,$}}

\def\ergps{\hbox{$\erg\s^{-1}\,$}}

\def\Jy{{\rm Jy}}
\def\kmps{\hbox{$\km\s^{-1}\,$}}

\documentstyle[psfig]{mn}

\title[The M87 Jet]
{The matter content of the jet in M87: evidence for an electron-positron jet}

\author[C.~S.~Reynolds, A.~C.~Fabian, A.~Celotti \& M.~J.~Rees]
{C.~S.~Reynolds$^1$, A.~C.~Fabian$^1$, A.~Celotti$^{1,2}$ and M.~J.~Rees$^1$\\
$^1$Institute of Astronomy, Madingley Road, Cambridge CB3 0HA\\
$^2$International School for Advanced Study, Trieste, Italy.}

\date{}

\begin{document}

\maketitle

\begin{abstract}
Recent observations have allowed the geometry and kinematics of the
M87 jet to be tightly constrained.  We combine these constraints with
historical Very Long Baseline Interferometry (VLBI) results and the
theory of synchrotron self-absorbed radio cores in order to
investigate the physical properties of the jet.  Our results strongly
suggest the jet to be dominated by an electron-positron (pair) plasma.
Although our conservative constraints cannot conclusively dismiss an
electron-proton plasma, the viability of this solution is extremely
vulnerable to further tightening of VLBI surface brightness limits.
The arguments presented, coupled with future high-resolution
multi-frequency VLBI studies of the jet core, will be able to firmly
distinguish these two possibilities.
\end{abstract}


\section{Introduction}

Jets are intimately linked with accretion processes.  They can be
found in almost all astrophysical situations in which matter is
believed to be undergoing disk accretion onto some central collapsed
object.  Scales range from the slow ($\sim 100$\kmps) sub-parsec jets
associated with forming stars to those found originating from powerful
active galactic nuclei (AGN) which are initially relativistic and can
propagate for hundreds of kpc before disrupting.  Despite intense
study, many basic questions remain open.  The underlying formation
mechanism is still uncertain (although hydromagnetic processes
associated with the accretion disk seem a promising candidate
mechanism: Blandford \& Payne 1982; Lynden-Bell 1995 and references
therein).  Moreover, whereas jets from forming stars are believed to
be molecular material entrained by a faster outflow of atomic
material, the basic nature of AGN jets has not been unambiguously
determined.  Possibilities include an electron-proton plasma
(i.e. normal material), an electron-positron plasma (i.e. a pair
plasma), Poynting flux or combinations of these.  This is an
observationally difficult issue to address: even the handful of well
studied jets only show power-law continuum radiation (presumably of
synchrotron origin) with no sign of any spectral features from
material moving with the bulk flow.  This is to be contrasted with the
Galactic source SS~433 in which the bulk flow of the jet is a known
source of atomic line emission.

Celotti \& Fabian (1993) have addressed the issue of the matter
content of jets in a sample of radio-loud quasars and radio galaxies.
Combining the synchrotron self-Compton constraints with indications of
the kinetic luminosity suggest that, for the sample as a whole, the
jets are either cold electron-positron flows or electron-proton flows
with an electron low-energy cutoff of $\sim 50\MeV$.  They argue for
the electron-proton case on the basis of annihilation constraints
under the assumption that the pairs originate from the inner regions
of the accretion flow.  However, the issue of the matter content of
any one given extragalactic jet (as opposed to a large, possibly
heterogeneous, sample) has not been addressed in detail.  An
(observationally based) determination of the matter content of an
individual well studied jet would be an important step in the study of
jet formation, propagation and emission.

M87 (NGC~4486) is a giant elliptical galaxy near the centre of the
Virgo cluster of galaxies.  It is associated with the radio source
(Virgo-A, 3C~274) and classified as a FR-I radio galaxy (Fanaroff \&
Riley 1974) on the basis of its edge-darkened morphology and low radio
luminosity ($P_{\rm 178MHz}\sim 1\times 10^{32}$\ergps\Hz$^{-1}$).
This galaxy contains the most spectacular example of an extragalactic
jet in the northern sky.  It was the first extragalactic jet to be
discovered (Curtis 1918) and, since then, it has been subjected to
intense observational studies at all wavelengths (see Biretta 1993 for
a recent review).  The complex knotty structure of the jet has been
resolved at wavelengths from the radio through to X-rays (Biretta,
Stern \& Harris 1991).  The spectrum and proper motions (Biretta, Zhou
\& Owen 1995) of the knots yield direct constraints on the physical
processes operating and the geometry of the system.  Hubble Space
Telescope (HST) studies of the base of the jet reveal a rotating gas
disk apparently lying normal to the jet direction (Ford et al. 1994;
Harms et al. 1994).  The projected appearance and velocity of the disk
suggests it to be at an inclination of $42\pm 5\degmark$ and to
enclose a mass of $(2.4\pm 0.7)\times 10^{9}\Msun$.  Also, Very Long
Baseline Interferometry (VLBI) has resolved structure within the core
of M87 on spatial scales of $\sim 0.01$\pc.  The VLBI studies will be
discussed in more detail in Section 3.1.  The synchrotron spectrum of
this jet has been recently examined by Meisenheimer, R{\"o}ser \&
Schl{\"o}telburg (1996).  This wealth of data makes M87 the most
promising candidate for any physical study of extragalactic jets.

In this paper we combine these observational constraints of the M87
jet with the theory of synchrotron radiation in order to constrain the
physical parameters of the jet.  In particular, we address the
particle density, magnetic flux density and the matter content of the
jet.  In order to keep this paper focussed, our treatment is as
model independent as possible.  The arguments we use depend only upon
an observationally well determined kinematic model and the standard
synchrotron jet model of Blandford \& K{\"o}nigl (1979).  Our
arguments are independent of any models (of which there are a
plethora) for the initial formation, collimination and acceleration of
the jet.  Section 2 briefly reviews the relevant observations and the
kinematic model to which they lead (Biretta 1993; Biretta, Zhou \&
Owen 1995).  Section 3 discusses the synchrotron self-absorption model
for the radio emission from the core and the subsequent constraints on
the particle density and the magnetic flux density.  Section 4 places
further constraints on the particle density of the jet using the large
scale limits on the total kinetic luminosity.  The resulting
constraints suggest the jet to be electron-positron dominated rather
than electron-proton dominated.  Section 5 discusses the robustness of
the result and some implications.  It is found that an electron-proton
jet possessing an electron population with a low-energy cutoff $\sim
10m_{\rm e}c^2$ also satisfies our constraints although this solution
is extremely vulnerable to further tightening of the VLBI surface
brightness lower limit.  Section 6 summarizes our conclusions.

Stellar surface brightness fluctuations give a distance to M87 of
$15.9\pm 0.9\Mpc$ (Tonry 1991), independently of Hubble's constant.
In the rest of this paper we assume a distance to M87 of $16\Mpc$.

\section{Kinematic model}

In this work, we use the kinematic model for the M87 jet presented in
Biretta (1993; hereafter B93) and Biretta, Zhou \& Owen (1995;
hereafter BZO).  This is based on detailed observations of proper
motions within the knotty structure of the jet.  In this model, the
inner jet is relativistic with a bulk flow Lorentz factor,
$\Gamma\approxgt 3$.  The prominent knot A marks the location of a
strong relativistic shock in which the flow slows from velocities of
$\sim c$ to $c/3$ (in the frame of reference in which the shock is
stationary).  In the observers frame, the Lorentz factor decreases
across the shock from $\Gamma\sim 3$ to $\Gamma\sim 1.5$.  The flow
suffers another shock at knot C at which it is decelerated to
subrelativistic velocities.  The jet then disrupts and merges into the
radio halo.  The orientation of the jet axis to the line of sight,
$\theta$, is between 30--40$\degmark$.

The relatively large angle to the line of sight (30--40$\degmark$)
puts severe limits on the extent to which relativistic beaming can be
important.  This fact will be extensively used in the later sections.
The effects of beaming are characterized by the {\it beaming
parameter}, $\delta$, defined by
\begin{equation}
\delta=\frac{1}{\Gamma(1-\beta {\rm cos}\theta)},
\end{equation}
where $\beta\,c$ is the bulk velocity of the material in the observers
frame.  For a given $\theta$, this has a maximum value of
\begin{equation}
\delta_{\rm max}=\frac{1}{{\rm sin}\theta},
\end{equation}
corresponding to $\beta={\rm cos}\theta$.  Evaluating for
$\theta=30\degmark$ gives $\delta_{\rm max}=2$ (and increasing
$\theta$ will result in decreasing $\delta_{\rm max}$).  Thus,
relativistic effects can only produce a moderate increase in the
observed luminosity of features moving along the jet.  Moreover, for
$\Gamma\approxgt 7$ (given $\theta=30\degmark$) relativistic
aberration diminishes the observed luminosity since it is beamed away
from the observer.

We note that relativistic beaming is important in rendering the
counter-jet undetectable.  Assuming the system to contain two
intrinsically symmetric (back-to-back) jets, a bulk Lorentz factor of
$\Gamma\approxgt 3$ and $\theta=30\degmark$ results in a
jet/counter-jet flux ratio of $\approxgt 300$.  The observational
limits on the jet/counter-jet (radio) flux ratio (based on the
non-detection of the counter-jet) are $\sim 150$ (Biretta 1993).

\section{Synchrotron self-absorption constraints}

\subsection{Basic model for the core emission}

Like many extragalactic radio sources, the core of M87 has a flat
radio spectrum (see Fig.~5 of Biretta, Stern \& Harris 1991; hereafter
BSH).  This is generally believed to be due to the superposition of
many synchrotron emitting components, each of which has a different
self-absorption frequency (Blandford \& K{\"o}nigl 1979).  In the
diverging flow of the inner jet, these components correspond to
emitting regions at different distances along the jet.  The observed
flux density at a given frequency is then dominated by the region of
the jet which becomes self-absorbed at that (Doppler shifted)
frequency.

Consider one such region of the jet which is just becoming
self-absorbed at the jet rest-frame frequency $\nu'_{\rm m}$.
Radiation with this frequency will suffer relativistic Doppler shifts
to an observer-frame frequency $\nu_{\rm m}=\nu'_{\rm m}\delta$.
Suppose the corresponding flux density from the self absorbed region
in the observers frame is $F_{\rm m}$.  Observations of the core at
the frequency $\nu_{\rm m}$ will measure this flux density plus
contributions from optically thin parts of the jet.  The total
observed core flux density at $\nu_{\rm m}$ will be $F_{\rm
tot}(\nu_{\rm m})=\eta F_{\rm m}$ where $\eta>1$ allows for the
emission from the optically thin parts of the jet.  Within the flat
spectrum regime, most of the emission must be dominated by the
self-absorbed region of the jet: thus we impose the condition that
$\eta\approxlt 2$.  If the observations also resolve the flat spectrum
core, the resulting dimension is a measure of the diameter of the jet
at the self-absorption point.

VLBI observations of M87 have been performed at a number of
frequencies.  Reid et al. (1989) present 1.66\,GHz VLBI observations
of the nuclear region of M87 showing a well collimated knotty jet
extending from an unresolved core.  These observations also show
subluminal motion within these knots suggesting that the pattern speed
is very much slower than the fluid speed.  The 5\,GHz observations of
Pauliny-Toth et al. (1981) just resolve the core to have a (major
axis) angular diameter of $\theta_{\rm d}=0.7$\,mas (corresponding to
0.06pc at the distance of M87) and a total core flux of $F_{\rm
m}=1.0\Jy$.  Higher frequency observations probe deeper into the core
of the jet before encountering the self-absorption surface as well as
achieving a better angular resolution.  Thus, 22\,GHz VLBI
observations with a resolution of $\sim 0.15$\,mas (corresponding to
0.01pc: Spencer \& Junor 1986; Junor \& Biretta 1995) reveal an
unresolved core (peak surface brightness 0.34\,Jy\,beam$^{-1}$) and a
stationary pattern of knots stretching for 4\,mas in the approximate
direction of the large scale jet.  The best resolution has been
achieved with 100\,GHz VLBI (B{\"a}{\"a}th et al. 1992) which reveals several
unresolved knots within 0.1\,mas of the core.  The flat spectrum
nature of the core is consistent with it being self absorbed up to
frequencies of at least 22\,GHz and possibly even 100\,GHz (Spencer,
private communication).

In the remainder of this section, we use the 5\,GHz VLBI constraints
on the flux and dimensions of the self-absorbed region to place
constraints on the relativistic (synchrotron emitting) proper particle
density, $n$, and magnetic flux density, $B$, at this location within
the jet.  We assume a uniform jet in the sense that $n$, $B$ and
$\beta$ are constant across cross-sections of the jet.

\subsection{Surface brightness of the self-absorbed region}

\begin{figure*}
\centerline{\psfig{figure=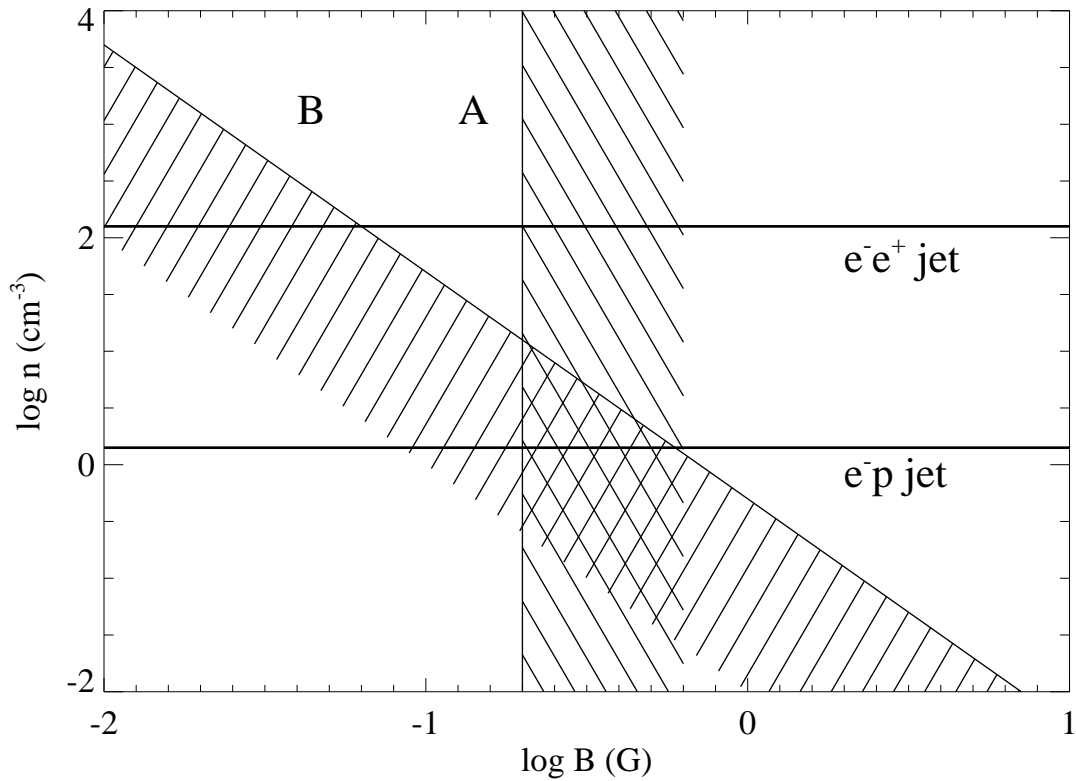,width=0.9\textwidth}}
\caption{Constraints on the $B$-$n$ plane imposed by synchrotron
self-absorption and total kinetic luminosity considerations.  The
$5\GHz$ VLBI surface brightness constraint on $B$ gives line A.  Line
B corresponds to the condition that resolved core of the jet is just
becoming optically thick to self-absorption.  The horizontal lines are
the densities derived by imposing $L_{\rm 43}=1$ for the two cases of
a pair jet (with $\gamma_{\rm min}=1$ and $\langle\gamma\rangle\approx
10$) and a cold jet of normal matter.}
\end{figure*}

\noindent Here we follow the work of Ghisellini et al. (1992; hereafter G92).
Suppose the relativistic electrons responsible for the synchrotron
radiation have a differential Lorentz-factor distribution of
$N(\gamma)=N_{\rm 0}\gamma^{-(2\alpha+1)}$ for $\gamma_{\rm
min}<\gamma<\gamma_{\rm max}$ where $\gamma$ is the electron Lorentz
factor as measured in the rest frame of the jet material.  

In the optically-thin regime (and well away from the low-frequency or
high-frequency cutoff), this produces synchrotron emission with a
spectral index $\alpha$.  For the M87 jet, $\alpha\sim 0.5$.  The
proper relativistic particle number density is given by
\begin{equation}
n=\int_{\gamma_{\rm min}}^{\gamma_{\rm max}}N(\gamma)d\gamma.
\end{equation}
For $\alpha>0$ and $\gamma_{\rm max}\gg\gamma_{\rm min}$, this
integral is dominated by the value of $\gamma_{\rm min}$.  The
particle number density, $n$, is then related to the normalization of
the relativistic electron distribution, $N_{\rm 0}$, by
\begin{equation}
N_{\rm 0}=2\alpha\gamma_{\rm min}^{2\alpha}n=\gamma_{\rm min}n,
\end{equation}
where we have used $\alpha=0.5$ in order to obtain the last
expression.  In this case the mean Lorentz factor of an electron is
given by
\begin{equation}
\langle\gamma\rangle=\gamma_{\rm min}\ln\left(\frac{\gamma_{\rm max}}{\gamma_{\rm min}}\right).
\end{equation}
For $\gamma_{\rm min}=1$, any reasonable choice of $\gamma_{\rm max}$
gives $\langle\gamma\rangle\sim 10$.

The synchrotron flux in the self-absorbed regime is independent of
particle density.  Thus, our constraints on $F_{\rm m}$ and
$\theta_{\rm d}$ can be translated into a constraint on $B$.  In
particular, there is a relationship between $B$ and the surface
brightness of the self-absorbed region (e.g. see G92) which gives,
\begin{equation}
B\approxlt 10^{-5}b(\alpha)\theta_{\rm d}^4F_{\rm m}^{-2}\nu_{\rm m}^5\frac{\delta_{\rm max}}{1+z}\,{\rm G},
\end{equation}
where $\theta_{\rm d}$ is in milliarcsec, $F_{\rm m}$ is in Jy and
$b(\alpha)$ is a function tabulated in G92.  The inequality results
from the limit on $\delta$.  From G92, $b(\alpha=0.5)\approx 3.2$, and
noting that $\delta_{\rm max}\approx 2$ gives $B<0.2\,{\rm G}$.  This
constraint is plotted on Fig.~1 (shaded areas represent forbidden
regions within the $n$,$B$ plane.)

\subsection{Optical depth of the self-absorbed region}

The point at which the jet is just becoming self-absorbed at an
observed frequency $\nu_{\rm m}$ is defined by $\kappa (\nu_{\rm m})
X=1$ where $\kappa(\nu)$ is the absorption coefficient for synchrotron
absorption at frequency $\nu$ and $X$ is the path length of the line
of sight through the jet.  Using the standard result for the
rest-frame absorption coefficient (Jones, O'Dell \& Stein 1974; Rybicki \& Lightman 1979)
and noting that $\nu_{\rm m}=\nu'_{\rm m}\delta$ gives
\begin{equation}
\kappa (\nu_{\rm m})=\frac{3^{\alpha+1}\pi^{1/2}g(p)e^2N_{\rm 0}}{8m_{\rm e}c}\nu_{\rm B}^{(3/2+\alpha)}\nu_{\rm m}^{-(5/2+\alpha)}\delta^{(5/2+\alpha)},
\end{equation}
(cgs units) where $\nu_{\rm B}$ is the cyclotron frequency,
$p=2\alpha+1$ and $g(p)$ is the given in terms of Gamma functions by
\begin{equation}
g(p)=\frac{\Gamma((3p+22)/12)\Gamma((3p+2)/12)\Gamma((p+6)/4)}{\Gamma((p+8)/4)}
\end{equation}
This expression for $\kappa(\nu)$ is valid for $\nu\gg\nu_{\rm
min}\approx\gamma_{\rm min}^2\nu_{\rm B}$, where $\nu_{\rm min}$ is
the low-energy cutoff in the synchrotron spectrum corresponding to the
low-energy cutoff in the electron energy distribution.  In this regime
the self-absorption depends only on the normalization of the
relativistic electron energy distribution and not on details of the
low-energy cutoff.  The spectrum of BSH shows no evidence for such a
low-energy cutoff thereby giving us confidence that the observed
frequencies lie in the domain of applicability of eqn (7).

We also need to relate the observer-frame line of sight path length
through the jet, $X$, with the physical jet diameter, $2r$.  Taking
into the account the relativistic transformations, the required
relation (for a cylindrical geometry) is
\begin{equation}
X=\frac{2r}{\delta\sin\theta}.
\end{equation}
Evaluating the above condition on the optical depth for $\alpha=0.5$
gives $nB^2>2\gamma_{\rm min}^{-1}\delta_{\rm max}^{-2}$ (cgs).  For
the moment we assume that $\gamma_{\rm min}=1$ (i.e. that the electron
energy distribution extends down to non-relativistic energies).
Relaxation of this assumption will be discussed in Section 5.3.  The
assumed geometry implies $\delta\approxlt 2$ leading to
$nB^2\approxgt0.5$ (cgs).  This is our second constraint on the
$n$,$B$ plane (Fig.~1).

\section{Kinetic luminosity constraints}

The total kinetic luminosity of the jet is an important observational
parameter.  In this section we estimate the total (time-averaged)
kinetic power of the jet by examining the energetics of the system as
a whole. We then assume the VLBI jet carries this same kinetic power
in order to place independent constraints on its properties.

\subsection{The global energy balance}

Ultimately, the kinetic energy carried by the jet can be radiated (at
the knots, for example), can perform work on the surrounding
interstellar medium (ISM) to expand the radio halo or can be stored in
relativistic particles within the radio halo.  Here we assess the
kinetic luminosity of the jet by summing these components.

\subsubsection{Radiated energy}

A multiwaveband study of the radiant emissions from the jet has been
performed by BSH.  By summing the observed emission of the jet
(dominated by emission from knots A and B) as tabulated in BSH, we
estimate that $\sim 3\times 10^{42}\ergps$ is radiated along the
length of the jet.  As stated above, relativistic beaming will not
have a major effect on the observed luminosity.

\subsubsection{Energy stored within the radio halo}

The large scale radio morphology of the system (Turland 1975) suggests
the presence of an inner radio halo ($\sim 5\kpc$ in extent) embedded
in a larger, more diffuse, halo ($\sim 50\kpc$ in extent).  The inner
structure is clearly associated with the current phase of nuclear
activity since the jet can be observationally traced from its origins
in the nucleus to the point at which it disrupts and feeds this radio
halo (e.g. see VLA images of Owen, Hardee \& Bignell 1980 and, more
recently, Hines, Owen \& Eilek 1989).  Spectral curvature arguments
suggest an age of $\sim 10^6\yr$ for this inner structure.  The
extensive outer halo is probably associated with a previous phase of
activity and will not be considered further here.  

Estimating the energy stored within the inner radio halo amounts to
determining the pressure of this structure.  Radio observations allow
the minimum (i.e. equipartition) pressure $p_{\rm min}$ of the
synchrotron emitting plasma to be determined.  The VLA data presented
by Hines, Owen \& Eilek (1989) show the inner radio lobe to have a
filamentary structure with the minimum pressure of the average
filament being $p_{\rm min}\sim 2$--$6\times 10^{-10}\dynpcmsq$.  The
brightest filaments have $p_{\rm min}\sim 10^{-9}\dynpcmsq$ whereas the
interfilamentary plasma has $p_{\rm min}\sim 4\times
10^{-11}\dynpcmsq$.  X-ray observations of thermal emission from the
interstellar medium (ISM) also allow the ISM pressure to be determined
at the vicinity where it comes into contact with the inner radio halo.
A deprojection analysis (Fabian et al. 1981) of data from the {\it
ROSAT} high-resolution imager (HRI) reveals an ISM pressure of $p_{\rm
X}\approx 5\times 10^{-10}\dynpcmsq$ (C.~B.~Peres, private
communication).  Thus the highest minimum pressures obtained from
radio observations exceed the ISM pressure by a factor of 2.  The
amount by which the radio halo is overpressured with respect to the
ISM can be estimated by considering the expansion of an overpressured
bubble.  Bicknell \& Begelman (1996) present such a model for the M87
system and show that the inner radio halo should be overpressured with
respect to the ISM by a factor of $\sim 3$.  Independent support for
this mild overpressure comes from the observed optical emission line
filaments discovered by Ford \& Butcher (1979) and studied in greater
detail by Sparks, Ford and Kinney (1993).  Bicknell \& Begelman (1996)
show how the expansion of this mildly overpressured halo energizes the
optical emission line filaments.  

To conclude this argument, the pressure of the inner radio halo is
likely to be $p\approx 1.5\times 10^{-9}\dynpcmsq$, leading to a
stored energy within the halo of $6\times 10^{56}\erg$.  The total
kinetic luminosity required to supply this energy in $10^6\yr$ is
$2\times 10^{43}\ergps$.  If there is a symmetric counter jet (as
assumed in the classic twin-jet model), the single-jet kinetic
luminosity which contributes to the relativistic particle energy in
the halo is $1\times 10^{43}\ergps$.

\subsubsection{Time-averaged kinetic luminosity, $L_{\rm K}$}

Summing the above contributions, the kinetic luminosity of the jet
(averaged over the duration of the current period of activity) is
$L_{\rm K}=10^{43}L_{\rm 43}\ergps$ where $L_{\rm 43}\sim 1$.  

An independent argument against a high-power jet can be put forward by
considering knot-A.  In the kinematic model of Biretta (1993), knot-A
represents a strong transverse shock in the flow where the jet
decelerates from $\Gamma\sim 3$ to $\Gamma\sim 1.5$.  A substantial
part of the kinetic luminosity would then be converted to other forms
at the shock.  Thus, the fact that we only see $\sim 2\times
10^{42}\ergps$ radiated at this knot suggests that the jet does not
possess a kinetic luminosity that is orders of magnitude above this
value.  Note that the fact that the flow remains collimated after
knot-A suggests that the kinetic energy is {\it not} transformed
purely into internal energy of the jet material.  Even in the more
physically plausible oblique shock models of Bicknell \& Begelman
(1996), a jet with a high kinetic luminosity ($L_{\rm K}\approxgt
10^{44}\ergps$) would be expected to radiate more than is observed.
Thus, a high jet power necessarily entails problems in understanding
the properties of knot-A.

Of course, the above arguments cannot rule out the possibility that
the jet has undergone a recent dramatic increase in kinetic
luminosity.  Such an increase could lead to a powerful VLBI core
whilst still keeping the total (average) energetics of the system at
the level observed.  If the jet has undergone a dramatic increase in
power, one might expect to observe the effects of this within the
inner jet (i.e. between the core and knot-A, where we have independent
arguments for a $L_{\rm K}\sim 10^{43}\ergps$ jet.)  However, the
inner jet is fairly featureless, displaying only the relatively weak
knots D, E, and F.  Furthermore, there is no precedent for the core of
M87 to undergo large amplitude (i.e. order of magnitude) variability.
Thus, to postulate a transient increase in jet power would be an
ad-hoc way to bypass the energetic constraints presented below.

\subsection{$L_{\rm K}$ and the physical jet parameters}

To relate the limits on $L_{\rm K}$ to physical parameters of the jet,
we assume the jet to be uniform in the sense that the density and
velocity are constant across a cross-section of the jet at a given
distance, $Z$, along the jet.  If positive charge carriers have mass
$m_{\rm +}$ per unit electron charge, the total jet kinetic luminosity
is given by
\begin{equation}
L_{\rm K}=\pi r(Z)^2\beta(\Gamma-1)\Gamma n(m_{\rm e}\langle\gamma\rangle+m_{\rm +}\langle\gamma_{\rm +}\rangle)c^3,
\end{equation}
where $r(Z)$ is the physical radius of the jet a distance $Z$ away
from its origin.  The inclusion of the $\langle\gamma\rangle$ and
$\langle\gamma_{\rm +}\rangle$ terms allow for energy that is advected
within the bulk flow for the electron and positive charge carrier
populations respectively.

From the above equation, $n$ can be expressed in terms of quantities
for which we have observational constraints and evaluated for the case
of an electron-proton jet ($m_{\rm +}=1836m_{\rm e}$) and an
electron-positron jet ($m_{\rm +}=m_{\rm e}$) at the 5\,GHz
self-absorption point.  We assume that $L_{43}=1$, $\Gamma=3$ and
$\langle\gamma\rangle=10$ (corresponding to the $\gamma_{\rm
min}\approx 1$ case).  Relaxation of these assumptions are discussed
in the next section.  For the electron-positron jet ($m_+=m_{\rm e}$,
$\langle\gamma_+\rangle=\langle\gamma\rangle\approx 10$) the result is
$n\approx 1.3\times 10^2\pcmcu$.  For the electron-proton jet, the
proton population will be essentially cold unless the average energy
per electron is comparable to or greater than the proton rest mass
(i.e. $\langle\gamma\rangle\approxgt 2000$).  In this case
($m_+=1836m_{\rm e}$, $\langle\gamma_+\rangle\approx 1$) we deduce
$n\approx 1.4\pcmcu$.  These results are shown as horizontal lines on
Fig.~1.

Examination of Fig.~1 strongly suggests the jet to be dominated by an
electron-positron plasma rather than an electron-proton plasma.  The
22\,GHz VLBI data and 100\,GHz VLBI data lead to a similar conclusion.
This is the main result of the present paper.

The only way for the above constraints to be compatible with an
electron-proton jet is if the relativistic electron population has a
low-energy cutoff at $\gamma_{\rm min}m_{\rm e}c^2$ with $\gamma\gg
1$.  Using either the 5\,GHz or 22\,GHz constraints requires
$\gamma_{\rm min}>10$.  The 100\,GHz constraints require $\gamma_{\rm
min}>100$ (assuming the core is still self-absorbed at this
frequency).  Physically, increasing $\gamma_{\rm min}$ allows the
normalization of the relativistic particle distribution, $N_{\rm 0}$,
to increase whilst holding the density, $n$, fixed.  Since the
synchrotron self-absorption constraints depend only on $N_{\rm 0}$,
varying $\gamma_{\rm min}$ allows the synchrotron constraints to
decouple from the kinetic energy constraints.  In particular, line B
on Fig.~1 will move downwards as $\gamma_{\rm min}$ increases such as
to keep the combination $n\gamma_{\rm min}B^2$ constant.
Observational and theoretical constraints on $\gamma_{\rm min}$ will
be considered in the next Section.  It will be seen that the range of
parameter space available for an electron-positron jet is relatively
small and not easily justifiable on physical grounds.  Furthermore,
the electron-proton solution is vulnerable to any future tightening of
these observational constraints.

\section{Discussion}

The above constraints argue for an electron-positron jet as opposed to
an electron-proton jet.  In this section we examine the robustness and
self-consistency of this result.  We then discuss some astrophysical
implications.

\subsection{The dimension of the self-absorbed core}

We have taken the angular size of the resolved core, as determined
from 5\,GHz VLBI, to represent the diameter of the jet at the point
where it just becomes self-absorbed.  More precisely, this should be
considered an {\it upper limit} to the jet dimension.  The VLBI core
dimension might plausibly represent the (projected) length of the
entire self-absorbed jet.  The effect of this uncertainty on our
constraints is simply addressed by re-calculating Fig.~1 for a smaller
jet diameter.  Suppose the physical jet diameter was a factor of 2
smaller than that inferred from VLBI observations.  The synchrotron
surface brightness constraint, eqn. (6), becomes tighter (i.e. the
upper limit on $B$ reduces) by a factor of $4$ (since the surface
brightness is twice that observed).  The optical depth constraint also
become tighter by a factor of 2.  The net effect is to raise the
(synchrotron) lower limit on $n\gamma_{\rm min}$ by a factor of $32$.
By constrast, the values of $n$ derived from the kinetic luminosity
will only be raised by a factor of 4.  Thus, the net effect is to
tighten the argument against an electron-proton jet.

If the jet in the core is, in fact, significantly narrower than the
VLBI resolution then even the electron-positron solution violates the
synchrotron constraints.  This is an argument against such a narrow
jet.

\subsection{The effect of a high velocity jet-core}

It is physically plausible that the jet starts as a high Lorentz
factor beam ($\Gamma\gg 10$) and subsequently decelerates via
dissipative processes or entrainment of surrounding material
(Begelman, Rees \& Sikora 1994; Sikora, Begelman \& Rees 1994).  The
VLBI jet may then possess a high Lorentz factor core surrounded by a
slower sheath ($\Gamma\sim {\rm few}$).  The emission from the core
would be largely beamed out of our line of sight and thus would be
swamped by emission from the slower sheath.  Our above synchrotron
constraints would apply only to the outer sheath since it would
dominate the observed radio emission.  However, since the kinetic
energy constraints are based on the large scale energetics of the
system, postulating the presence of an unseen rapid core reduces the
inferred kinetic luminosity of the outer sheath and thus makes the
case against an electron-proton jet stronger.

\subsection{Synchrotron self-Compton radiation and observational limits on $\gamma_{\rm min}$}

\subsubsection{Synchrotron self-Compton radiation}

The production of high-energy synchrotron self-Compton (SSC) radiation
is an unavoidable consequence of the synchrotron process.  We must
check that our jet models are consistent with the constraint that the
SSC radiation does not violate the observed X-ray limits on the core
flux.  We can also use this constraint to place observational limits
on $\gamma_{\rm min}$.

The SSC X-ray flux density, $F_{\rm X}$ from the self-absorbed region
(size $r$) is given by G92 as
\begin{equation}
F_{\rm X}=\frac{2\alpha F_{\rm m}n\sigma_{\rm T}r\gamma_{\rm
min}^{2\alpha}}{t(\alpha)}\left(\frac{\nu_{\rm m}}{\nu_{\rm
X}}\right)^\alpha\ln\left(\frac{\nu_{\rm b}}{\nu_{\rm m}}\right),
\end{equation}
where $\nu_{\rm X}$ is the X-ray frequency, $\nu_{\rm b}$ is the
high-energy cutoff in the synchrotron spectrum and $t(\alpha)$ is a
function tabulated in G92.  We typically assume that $\nu_{\rm b}$
lies somewhere between optical and X-ray frequencies.  Evaluating this
flux density at $1\keV$ for parameters relevant to the 5\,GHz self
absorbed region gives
\begin{equation}
F_{\rm X}\approx 3.0\times 10^{-11}n\gamma_{\min}\ln\left(\nu_{\rm
b}/\nu_{\rm m}\right)\Jy.
\end{equation}
where
\begin{equation}
\ln(\nu_{\rm b}/\nu_{\rm m})\sim 10     
\end{equation}
for physically plausible values of $\nu_{\rm b}$.  For consistency,
this must be less than the observed X-ray core flux density of $\sim
3.5\times 10^{-7}\Jy$ (BSH).  We note that using the result of Section
3.3 we can rewrite this as a function of the VLBI core flux density
and the magnetic field $B$ only.  This gives a lower limit on the
magnetic field of $B\approxgt 0.01$\,G.

For pair jets, we have shown from our kinetic luminosity arguments
that $n\approx 10^2\pcmcu$.  The production of synchrotron
self-Compton radiation is then consistent with observations provided
$\gamma_{\rm min}\approxlt 10$ (in which case, a substantial fraction
of the observed X-rays will originate from the SSC process).  For an
electron-proton jet ($n\approx 1\pcmcu$) the corresponding limit is
$\gamma_{\min}\approxlt 10^3$.

\subsubsection{SSC limits and $\gamma_{\rm min}$}

The SSC limits give $B\approxgt 0.01\,{\rm G}$.  Thus, an electron
energy distribution that cuts off at $\gamma_{\rm min}$ will produce a
low-energy cutoff in the synchrotron spectrum at frequency $\nu_{\rm
c}\sim\gamma_{\rm min}^2(B/{\rm G})\MHz$.  Observationally, $\nu_{\rm
c}<1\GHz$ leading to the condition that $\gamma_{\rm min}<100$.  Thus,
if $10<\gamma_{\rm min}<100$, the electron-proton jet is consistent
with our 5\,GHz and 22\,GHz constraints.  If the core emission is
still dominated by self-absorbed radiation at 100\,GHz, the
corresponding VLBI constraints are incompatible with an
electron-proton jet since compatibility would demand $\gamma_{\rm
min}\sim 100$ which would lead to an unobserved low-frequency cut off
at $\nu_{\rm c}\sim 100\GHz$ (note that the SSC limit on the magnetic
field at the 100\,GHz self-absorption location is $B\approxgt 10$\,G).

\subsection{Theoretical issues related to $\gamma_{\rm min}$}

In the absence of further observational limits on $\gamma_{\rm min}$,
we briefly discuss some related theoretical issues.

Let us suppose that the relativistic electron population at the 5\,GHz
self-absorbed point did, in fact, possess a low-energy cutoff
corresponding to $\gamma_{\rm min}\sim 10$--$100$.  Such an energy
distribution cannot be a relic of the conditions imposed near the base
of the jet: synchrotron and inverse Compton losses would have cooled a
significant number of these electron to energies below $10m_{\rm
e}c^2$.  Thus, some in-situ physical mechanism must be maintaining
this low-energy cutoff (i.e. by either reheating the
electrons/positrons or injecting fresh high-energy pairs).

One mechanism that has been proposed for injecting high-energy pairs
into a jet is that of high-energy hadronic (primarily proton-proton,
hereafter p-p) collisions (Falcke \& Biermann 1995; Biermann, Strom \&
Falcke 1995).  Such collisions produce both charged and neutral pions.
The charged pions then decay into electrons or positrons,
corresponding to the $\pi^-$ and $\pi^+$ cases respectively, with an
initially high energy ($\gamma\sim 100$) due to the large rest mass of
the pions.  However, the p-p interaction cross section is
approximately $\sigma_{\rm pp}\approx \sigma_{\rm T}/10$ (and
decreases as $\gamma$ increases).  Thus, the optical depth to p-p
collisions at the location where our VLBI constraints are applied is
$\approxlt 10^{-8}$ (where we have used the fact that the number
density of protons is constrained to be $n_{\rm p}\approxlt 1\pcmcu$
by the kinetic luminosity constraint).  We conclude that this
mechanism is not relevant in this case.  In fact, it would only be
relevant at the base of very powerful (and compact) jets.  We note
that hadronic processes may be of importance in compact plasmas near
the centre of the accretion flow (Sikora et al. 1987; Begelman, Rudak
\& Sikora 1990).  In the absence of any other mechanism that inject
high-energy electrons, any low-energy cutoff seen at the 5\,GHz
self-absorption point is not likely to be due to such injection.

Various heating mechanisms that could lead to $\gamma_{\rm min}>1$
have been discussed by G92.  The most likely heating mechanism is that
due to synchrotron self-absorption (the so-called `synchrotron
boiler'; Ghisellini, Guilbert \& Svensson 1988).  For the region under
consideration here, electrons radiating at the self-absorpion
frequency (5\,GHz) have $\gamma\sim 100$.  If only synchrotron
processes were important (i.e. the plasma is magnetically dominated),
the electron energy distribution below this would possess a
quasi-thermal component (due to the efficient excahnge of energy
between the relativistic electrons).  The effective low-energy cutoff
would then lie somewhere below $\gamma_{\rm min}\sim 100$.  However,
if inverse Compton losses are important (as is probably the case for jets
on the VLBI scale), the effective low-energy cutoff will be much
reduced and, indeed, synchrotron heating can become irrelevant.  A
detailed treatment of this phenomenon is beyond the scope of this
paper.

\subsection{Annihilation of the pairs}

The cross-section for annihilation of cold (i.e. non-relativistic)
pairs is given by $\sigma_{\rm a}=(3/8)\sigma_{\rm T}$ where
$\sigma_{\rm T}$ is the Thomson cross-section.  [As the pairs become
relativistic, the cross-section declines in a manner analogous to the
Klein-Nishina decline of the Thomson cross-section.]  In the comoving
frame, the rate of annihilation of cold pairs is then
\begin{equation}
{\dot n}=\frac{3}{8}c\sigma_{\rm T}n^2,
\end{equation}
leading to an annihilation timescale of
\begin{equation}
t_{\rm ann}=\frac{8}{3c\sigma_{\rm T}n}.
\end{equation}
In the outer, low-density parts of the jet this timescale is long and
annihilation is not an important process.  However, in the higher
densities found near the jet core the annihilation rate increases.  It
is instructive to estimate the distance from the nucleus at which the
annihilation timescale, $t_{\rm ann}$ equals the flow timescale
$t_{\rm flow}=Z/\Gamma c$ where $Z$ is the distance from the core.  If
the jet has a constant opening angle and constant velocity so that the
density falls off as $n\propto Z^{-2}$, equating $t_{\rm ann}$ and
$t_{\rm flow}$ gives the {\it annihilation radius}
\begin{equation}
Z_{\rm ann}=\frac{3\sigma_{\rm T}n_{\rm 0}Z_{\rm 0}^2}{8\Gamma},
\end{equation}
where $n_{\rm 0}$ is the particle density at a distance $Z_{\rm 0}$
along the jet.  Our (cold-jet) kinetic luminosity constraints give
$nr^2=4\times 10^{36}\cm^{-1}$.  If we assume a half opening angle of
$\approxgt 0.1$\,rad (such as to agree with the opening angle on
larger scales) then $Z\approxlt 10r$. Together with $\Gamma=3$, this
evaluates to give $Z=3\approxlt 10^{13}\cm$.  This is to be compared
with the Schwarzschild radius of $R_{\rm s}\approx 1\times 10^{15}\cm$
($M=3\times 10^9\Msun$ assumed).  Thus it is feasible for the jet to
have originated from a pair plasma formed at the centre of the
accretion flow.

\subsection{The FR-I/FR-II dichotomy}

Celotti \& Fabian (1993) utilized SSC constraints on a sample of radio
galaxies and radio-loud quasars (dominated by FR-II type objects) and
argued for either an electron-positron jet with $\gamma_{\rm min}\sim
1$ or an electron-proton jet with a $\gamma_{\rm min}\sim 100$.  For
these powerful sources, it was found that the annihilation radius lies
well outside the region occupied by any compact pair plasma associated
with the inner accretion flow.  Thus it is not possible to have the
pair jet flowing freely from the immediate vicinity of the black hole.
This argument was used to favour the electron-proton case.  However,
we have presented arguments for an electron-positron jet in M87 (a
classical FR-I source).  These results suggest one of two
possibilities:

a) FR-I and FR-II may possess jets of an intrinsically different
nature, at least by the time the jet has propagated to VLBI scales:
i.e. FR-I sources may possess electron-positron VLBI jets whereas
FR-II sources may possess VLBI electron-proton jets.  The
morphological differences between FR-I/FR-II sources may reflect this
underlying difference in the jet (for example, via the ease of
collimation).  However, it is difficult to envisage the physical
processes leading to such a dichotomy and how they could be related to
the source power.

b) The annihilation radius constraint does not apply and both classes
of sources possess electron-positron jets with $\gamma_{\rm min}\sim
1$.  In this case, the jet would have to propagate from the vicinity
black hole out to at least the annihilation radius in some other form
such as a Poynting flux dominated flow.  Beyond the annihilation
radius, electromagnetic cascades can lead to copious pair production
within the Poynting flux dominated flow.

\section{Conclusions and definitive observations}

The M87 jet has the best determined geometry and kinematics of any
extragalactic jet in the sky.  M87 is also one of the very few
galaxies in which the mass of the central black hole ($2.4\pm
0.7\times 10^9\Msun$) is well constrained (by HST observations).  This
makes M87 an obvious system in which to study the physical properties
of extragalactic jets.

We have utilized the standard theory of synchrotron self-absorbed
cores in order to constrain the magnetic field, $B$, and the
relativistic (proper) particle density, $n$, of the jet.  The 5\,GHz
data implies $n\gamma_{\rm min}\approxgt 10\pcmcu$.  For this jet to
carry the kinetic liminosity inferred from global energetic arguments,
the density is either $n\sim 1\pcmcu$ (electron-proton jet) or $n\sim
10^2\pcmcu$ (electron-positron jet).  We cite this as evidence for an
electron-positron dominated jet rather than an electron-proton
dominated jet.  An electron-proton jet with $10<\gamma_{\rm min}<100$
is consistent with the present (conservative) constraints but is
extremely vulnerable to further tightening of the VLBI limits.  If the
core seen in 100\,GHz VLBI data is self-absorbed, the electron-proton
jet is ruled out.  Some theoretical arguments are also presented
detailing the problems of the high-$\gamma_{\rm min}$ scenario.

To make further observational progress, the self-absorbed core has to
be identified and resolved.  This requires high-resolution
multi-frequency VLBI observations.  The sub-arcsec X-ray imaging
capability of {\it AXAF} will separate the core X-ray emission from
other components (e.g. inner knot emission) and thus allow stricter
limits to be placed on the SSC radiation of the self-absorbed core.
Such data, together with arguments similar to those presented here,
should allow the matter content of this archetypal extragalactic jet
to be unambiguously determined.

\section*{Acknowledgements}

CSR thanks PPARC for support.  ACF, AC and MJR thank the Royal Society
for support.  We are grateful to Ralph Spencer for his useful comments
on the interpretation of the VLBI data.

\end{document}